# From Crowdsourcing to Crowdmining: Using Implicit Human Intelligence for Better Understanding of Crowdsourced Data

Bin Guo, *Senior Member*, Huihui Chen, Yan Liu, Chao Chen, *Member*, Qi Han, *Member*, Zhiwen Yu, *Senior Member*

*Abstract*—With the development of mobile social networks, more and more crowdsourced data are generated on the Web or collected from real-world sensing. The fragment, heterogeneous, and noisy nature of online/offline crowdsourced data, however, makes it difficult to be understood. Traditional content-based analyzing methods suffer from potential issues such as computational intensiveness and poor performance. To address them, this paper presents CrowdMining. In particular, we observe that the knowledge hidden in the process of data generation, regarding individual/crowd behavior patterns (e.g., mobility patterns, community contexts such as social ties and structure) and crowd-object interaction patterns (flickering or tweeting patterns) are neglected in crowdsourced data mining. Therefore, a novel approach that leverages implicit human intelligence (implicit HI) for crowdsourced data mining and understanding is proposed. Two studies titled CrowdEvent and CrowdRoute are presented to showcase its usage, where implicit HIs are extracted either from online or offline crowdsourced data. A generic model for CrowdMining is further proposed based on a set of existing studies. Experiments based on real-world datasets demonstrate the effectiveness of CrowdMining.

## I. Introduction

CROWDSOURCING system enlists a number of humans to help solve a wide variety of problems with their skills, experiences, and intelligence. The first trend of such systems appeared on the World-Wide-Web [1], with the purpose of building artifacts (knowledge bases, maps) and performing tasks (e.g., Amazon Mturk[1], Crowdflowers[2]). In the Web 2.0 era, the rapid development of social media services (e.g., Twitter[3]) makes it easier for the crowd to share contents, feelings, and knowledge. This has become another way to present crowd intelligence (e.g., using Twitter data to detect and characterize events [2]).

Recently, with the prevalence of sensor-equipped mobile devices, we have envisioned a new trend of crowdsourcing systems that often work in the real-world settings, i.e., the so-called mobile crowdsourcing (MCS) [3], [4]. Specifically, the participants in MCS use their mobile devices to perform large-scale sensing tasks, such as collecting traffic information. In contrast with traditional sensor networks, MCS aims to leverage the power of average users and their associated mobile devices to achieve large-scale sensing. Representative examples on MCS include FlierMeet [5], SmartPhoto [6], Urban Resolution [7], and so on.

For either online or mobile crowdsourcing tasks, data collection and sharing is a primary task [8]. The ever increasing participants contribute large volume of data [9]. The data contain rich and complex information, which can be used in a variety of applications such as event reporting [2], traffic dynamics [10], disaster management [6], pollution monitoring [7], and object imagery [11]. The crowdsourced data can hardly be used directly to yield usable information [12]. Intelligently analyzing and processing crowdsourced information can help prepare data to maximize the usable information, thus returning the benefit to the crowd. The major features and relevant issues are characterized below.

- *Noisy*. The data contributed by average users vary in quality and reliability. Some people can contribute accurate information while others may not. Data from distributed 'human sensors' are often redundant, e.g., people may Tweet similar posts, pictures taken nearby can be highly-duplicate [5].
- *Heterogeneous*. Crowd contributed data often contain rich yet heterogeneous information, in the forms such as texts, images, and audio/video clips. Varied interaction information (e.g., reposts, likes) is also available with the development of Web 2.0 services. With the prevalence of sensing-equipped devices, we envision more and more sensory information collected in crowdsourcing systems, such as locations and activities.
- *Fragmented*. The data are fragmented yet correlated regarding latent objects or themes, e.g., places, events, products, and humans. For example, people can express their opinions about the different aspects of a product (e.g., iPhone 6). They can also take pictures of different stages of an event (e.g., a street performance).

The above features make it challenging to analyze and

This work was partially supported by the National Key R&D Program of China(2017YFB1001803), National Basic Research Program of China (No.2015CB352400), and the National Natural Science Foundation of China (No. 61772428, 61725205).

B. Guo, H. Chen, Y. Liu, and Z. Yu are with School of Computer Science, Northwestern Polytechnical University, Xi'an 710072 China (e-mail: guob@nwpu.edu.cn).

C. Chen is with the Department of Computer Science, Chongqing University, 400044 China (e-mail: cschaochen@cqu.edu.cn). Q. Han is with Colorado School of Mines, US (e-mail: qhan@mines.edu).

[1] http://www.mturk.com/
[2] http:// www.crowdflower.com/
[3] http://www.twitter.com/

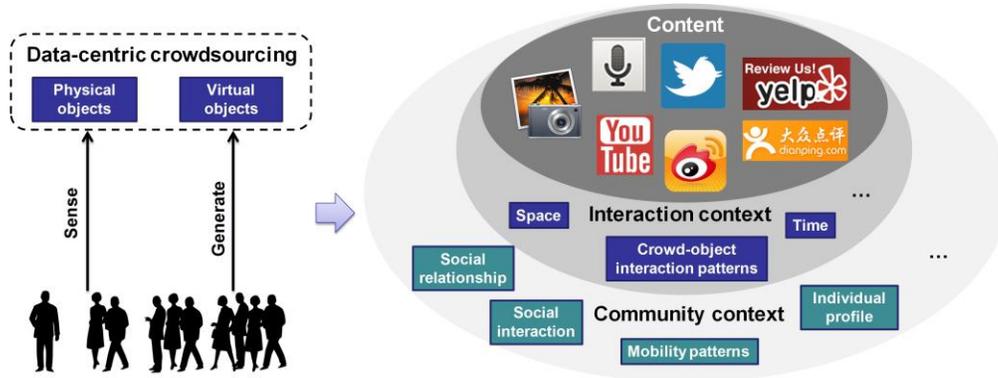

Figure 1. Human and data-centric crowdsourcing: a deep insight.

understand crowd-contributed data. Existing methods towards this are mostly based on the data itself [13], [14]. Analyzing the content of huge volume of data is usually computationally intensive and thus works poorly in many cases. For many problems such as image understanding, humans can still perform more accurately and efficiently than a machine. To this end, some researchers have tried to involve explicit human intelligence (HI) to support crowdsourced data mining, for the tasks such as data classification [15] and query-answering [16]. Though effective, active participation is often boring for the participants. We aim to address the challenge from a new perspective: *harnessing the implicit HI to better understand crowdsourced data*. We notice that the knowledge hidden in the process of data generation, regarding individual or crowd behavior patterns are neglected in crowdsourced data mining [17]. There are some examples of implicit HI, including online tweeting or flickring (picture shooting & sharing) patterns, human mobility patterns, social interaction behaviors, etc. Such knowledge is implicit and does not need active response from people. This paper studies how to measure and use the aggregated effects of crowd behavior patterns for crowdsourced data mining (i.e., CrowdMining).

We have made the following contributions.

(1) Giving an insight of data-centric crowdsourcing and characterizing the difference between explicit and implicit HI in CrowdMining.

(2) Proposing a formal model for crowdsourced data understanding with implicit HI, regarding general data mining tasks such as filtering, classification, and grouping.

(3) Presenting two case studies to demonstrate how implicit HI can be used for crowdsourced data understanding. More existing studies are also summarized.

(4) Experiments over real-world datasets to the two case studies demonstrate the effectiveness of using implicit HI.

The remaining paper is organized as follows. Section II gives a characterization of CrowdMining, including its layered data structure and the scope of implicit HIs. Section III presents a generic CrowdMining model. Two case studies are presented in Section IV and V to demonstrate the usefulness of using implicit HI. Finally, Section VI gives the conclusion of this paper and the vision of future research directions.

## II. CHARACTERIZING CROWDMINING

We first give an anatomy of data-centric crowdsourcing. A comparison of explicit and implicit HI for crowdsourced data understanding is then presented.

### A. A Deep Insight into Data-Centric Crowdsourcing

The core concept in crowdsourcing has been around for some time, and it is broad and incorporates a wide range of apps. This paper focuses on some specific tasks, named data-centric crowdsourcing, where crowdsourced data is contributed by people on the mobile app. Early practices are user-generated content in online social media, such as Wikipedia[4], Twitter, Yahoo Answers[5], Yelp[6]. For example, Twitter has been proven to be a useful crowdsourcing tool, for the tasks such as work collaboration, collective wisdom, and emergent event identification [1]. Recently we have witnessed the development of mobile crowdsensing by capturing the dynamics of real-world objects (e.g., a place [18], an event [19], and a shop [20]). Overall, data-centric crowdsourcing is about human-object interaction/association, where people generate data about virtual objects in social media or sensory data about physical objects in the real world, as shown in Fig. 1.

A deep insight of crowdsourced data has three corresponding layers: content, interaction context, and community context.

- *Content* refers to user-contributed data, i.e., real-world sensing data or user-generated content in social media.
- *Interaction context*. It refers to the relationship between human and data, i.e., how data is generated or contributed by human.
- *Community context*. For a selected crowdsourced dataset, there will be an associated community that participates in data contribution. The information about the community and its members (i.e., community contexts), such as individual profiles, user preferences, social ties, interaction dynamics, and behavior patterns, are important information to understand crowdsourced data.

---

[4] http://www.wikipedia.org/
[5] http://answers.yahoo.com/
[6] http://www.yelp.com/

## B. Explicit and Implicit Human Intelligence

We present and make a comparison of the three concepts related to 'intelligence' in crowdsourcing.

*Machine intelligence* (*MI*) is the intelligence exhibited by machines [21], i.e., the so-called artificial intelligence. For data-centric crowdsourcing, the approaches used mainly include image recognition, natural language processing, statistical learning, and reasoning. It is motivated by enabling machines to simulate *human intelligence* (*HI*) [22]. In this paper, we characterize HI into two types: *explicit HI* and *implicit HI*. The difference between them is about the awareness of tasks on processing crowdsourced data.

For *explicit HI*, people are aware of the data processing/understanding tasks and they actively perform the tasks using their abilities and knowledge. There have been numerous works in this area. For example, Digg[7] employs people to assign tags to articles (i.e., article classification). ReCAPTCHA [23] uses human intelligence for irregular character recognition, which is hard to be achieved by MI. Hashtags in Twitter help cluster the tweets with the same topic. Links in Wikipedia identify the association among the items. Flock [24] employs the crowd to suggest predictive features for data labeling, and the suggested features are then weighted and used in a machine learning model. We list the typical abilities relevant to explicit HI in Fig. 2. Explicit HI can be directly used (e.g., ReCAPTCHA, Hashtags in Twitter) or integrated with MI (e.g., Flock [24]) for crowdsourced data understanding.

For *implicit HI*, people are not aware of the crowd mining tasks and they do not actively contribute their intelligence and knowledge for data processing/understanding. More specifically, such knowledge is implicit, which is hidden in the process of data generation, and does not need active response from people. Alternatively, we leverage human-relevant knowledge implied in crowdsourced data contribution for data understanding, such as online tweeting or flickring, human mobility patterns, etc. In data-centric crowdsourcing, it refers to the two human-relevant layers in Fig. 1, i.e., interaction and community contexts [25]. Various types of human intelligence are embedded in the data sensing or content generation process.

- *Interaction contexts* are based on human perception, decision making, opinions, etc.
- *Community contexts* are related to individual traits, community structure, and social/individual behavior patterns.

They are often used indirectly for crowdsourced data understanding, i.e., used as features or parameter inputs of MI. We list the often-used implicit HI in Fig. 2. There are various data mining tasks that can be performed by means of crowdsourcing, such as data filtering and classification.

## III. TOWARDS A GENERIC CROWDMINING MODEL

We make a brief review of other recent studies where implicit HI is also partly used for crowdsourced data understanding. Afterwards, we propose a generic model and framework for building crowd mining systems.

[7] http://www.digg.com

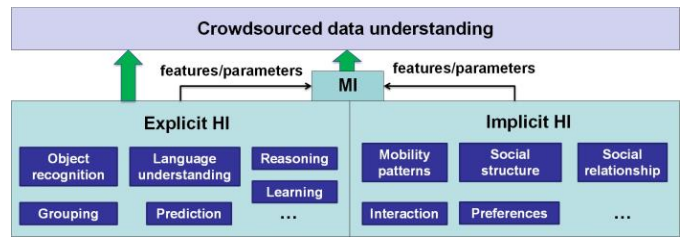

Figure 2. Explicit and implicit HI.

### A. Other data-centric crowdsourcing tasks

We present the related studies from two aspects: mobile crowdsourcing and online crowdsourcing.

*(1) Mobile crowdsourcing.* FlierMeet is an MCS app for public information (distributed fliers in the city) sensing and tagging. There are two aspects that implicit HI is used, details can be found in our previous work [5]. *a) Best shooting angle detection.* Crowd-contributed flier pictures are redundant, presenting the opportunity to select high-quality data. One of the issues is how to select pictures with the best shooting angle. Pictures shot in the frontal direction normally have better quality than yawed ones. To determine the best-shooting direction for the fliers on a certain bulletin board, a crowd-powered approach is proposed: we measure the central tendency of the collected heading angles to the board, and use the central value as the best shooting angle. *b) Flier tagging with crowd-flier interaction patterns.* The tags (popular, hot, community-specific) learned are useful for flier sharing and personalized information suggestion. Specifically, we utilize various contexts (e.g., spatio-temporal info, flier posting behaviors) to group similar reposts. We further identify a novel set of crowd-object interaction hints (e.g., crowd-flier entropy, temporal feature, number of reposts) to predict the semantic tags of reposts.

CrowdMap [26] leveraged aggregated user motion traces to generate indoor floor plan and room shape. The premise is that users would be able to move across all edges and corners in an indoor environment. Crowd visiting patterns at the target places are leveraged as implicit HI for outlier filtering and key point detection. Movi [19] used group behavior patterns (laughter, group rotation, shared viewing, ambience fluctuation) to identify potential interesting scenes from crowdsourced video clips in social events. Data contributed by the crowd may be semantically or visually relevant, to eliminate data redundancy and reduce network overhead, PicPick [27] proposed a Pyramid Tree approach to select an optimal set of pictures from picture streams based on multi-dimensional task constraints (e.g. spatio-temporal contexts, single or multiple shooting angles). Besides event localization, we have also investigated event picture selection and multi-grained summarization using crowd photographing entropy features [28].

*(2) Online crowdsourcing.* There have also been many studies of online crowdsourced data understanding with implicit HI. Cranshaw *et al.* [29] examined crowdsourced location traces in a location sharing social network for inter-user friendship prediction. A set of location-based features are introduced for analyzing the diversity of a geographic region (e.g., a shopping mall, a restaurant), including visitor counting, visiting frequency, location entropy (measuring the diversity of unique visitors of a location), and so on. Crowd mobility patterns (implicit HI) are thus used to predict the social relationship between users.

Redi *et al.* [30] used visual cues extracted from the profile pictures in FourSquare to guess the ambiance (e.g., calm, relaxing, reading, and cramp) of places. The visual cues are human-oriented, including aesthetics, emotions, demographics, self-presentation, etc. For example, dark pictures (those lacking brightness) are indeed used by people who go to cramped places (aesthetics); people going to strange places do not smile, while those going to places catered to attractive people do so (emotion). Facebook likes and FourSquare check-ins have been found useful for predicting personal traits and attributes [31-33]. TwitInfo [34] leveraged the crowd posting patterns (e.g., peaks) in Twitter for subevent detection in sports. Lin et al. [35] selected representative data from numerous microblogging posts for event summary, leveraging both content and posting context features.

### B. The Generic Model

As discussed above, understanding of crowdsourced data is useful for at least the following types of applications.

- *Object profiling*. Characterizing the features of an object, such as a place, a shop, or a flier.
- *Human profiling*. Characterizing the features of a person or social relations.
- *Event sensing*. Detecting, segmenting, and summarizing events.

Table 1. Understanding of crowdsourced data with implicit HI.

| Task type | Related work | The usage of Implicit HI |
|---|---|---|
| Filtering | Data quality measurement [5], Eliminating errors and outliers [26] | Aggregated shooting behaviors, Aggregated mobility traces |
| Classification & Tagging | Flier tagging [5], Friendship prediction [29], Human trait understanding [31-33], Place categorization [30] | Group structure, Crowd-object interaction patterns, Mobility patterns |
| Clustering & Segmentation | Highlight detection [19], Subevent detection [34] | Group behavior patterns (rotation, laughing), Crowd posting patterns |
| Data selection | Redundancy elimination [27], Event summary [35] | Picture shooting contexts, Posting patterns |

The usage of implicit HI for data processing incorporates at least the following task types, and we make a summary shown in Table 1.

- *Data filtering*. Filtering noisy or low-quality data.
- *Data classification and tagging*. Categorizing the data or assigning tags to the data.
- *Data clustering & segmentation*. Grouping redundant data. For evolutionary sensing objects such as events, it is often important to segment the data stream.
- *Data selection*. Selecting representative data from the redundant data set.

### C. The Framework

To facilitate the development of CrowdMining applications, a generic system framework is essential. We have proposed a conceptual framework for CrowdMining systems, as shown in Fig. 3. It can be a starting point to build CrowdMining applications with framework support.

The framework consists of the following components: The *data crowdsourcing* layer is responsible for collecting data from participatory sensing communities or online social media. The *implicit HI* layer applies diverse machine processing techniques to extract implicit HI, such as human behavior patterns and mobility patterns. The *crowdsourced mining* layer leverages the learned implicit HI to various data mining tasks, such as data filtering, data selection, and clustering. Finally, the *application layer* includes a variety of potential services that can be enabled by the availability of CrowdMining.

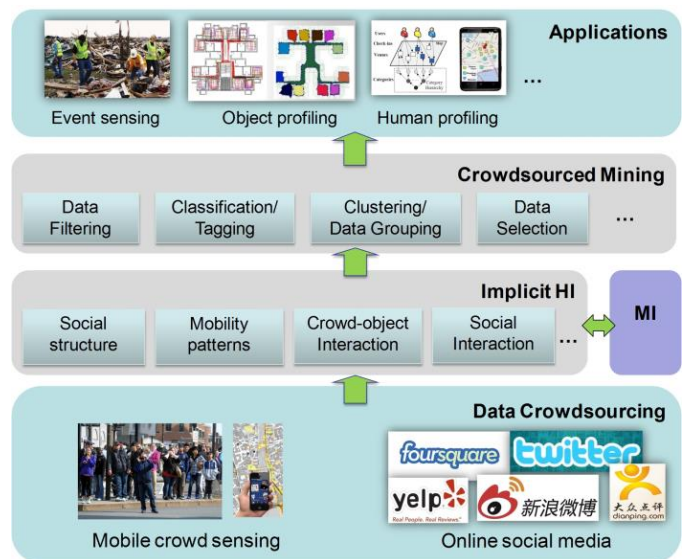

Figure 3. A generic framework for CrowdMining.

## IV. CROWDMINING ON THE WAY WITH CASE STUDY-1: CROWDEVENT

We present two of our studies to demonstrate the power of using implicit HI for crowdsourced data understanding in Sections III and IV. For each case study, we first describe a scenario and discuss the issues related to data processing, and then present our methods that leverage implicit HI. Experiments to validate the effectiveness of these methods will finally be presented.

### A. The Scenario and Data-specific Tasks

Figure 4 shows that four people (i.e., reporter *A~D*) take pictures of a street performance. Absent people may want to watch it. A real-time mobile visual sensing and sharing system can facilitate sharing this event with these absentees.

There are many challenges to be addressed in this use case. Here we mainly focus on two of the following issues.

*a) Event localization*. There might be several co-occurring events nearby. They can be distinguished by their location. However, the locations of picture-takers are not necessarily the location of events (the distance between them can range from several meters to tens of meters), so simply using GPS coordinates of the event reporter as the event location could be misleading. We thus should find an approach for event localization.

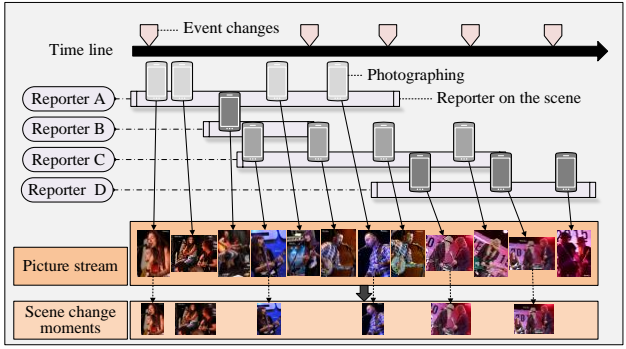

Figure 4. The CrowdEvent scenario.

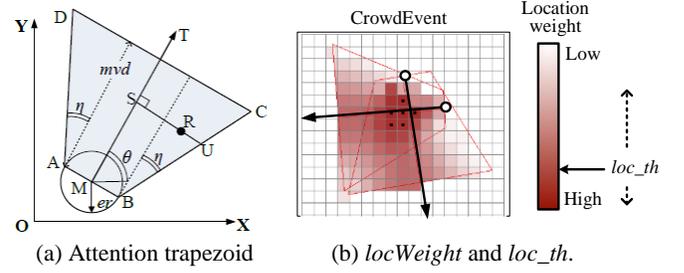

(a) Attention trapezoid    (b) *locWeight* and *loc_th*.

Figure 5. Event localization with shared attention.

*b) Sub-event detection*. Events usually evolve and can be divided into a series of sub-events. To make better visualization of events to readers, it is necessary to find an efficient method to segment the crowdsourced event picture stream into sub-events. Existing studies concentrate on using visual content similarity to segment events [36-39]. They suffer from issues such as semantically incorrect segmentation (e.g., two pictures with similar content may represent different event stages) and high computational overhead.

### B. The Usage of Implicit HI

We develop CrowdEvent and use implicit HI to address the two issues above.

*a) Event localization with shared attention*. Photographing contexts, such as the location and shooting angle of a viewer, can represent her attention area (e.g., the event area) in picture taking. However, a person often takes pictures from one direction and thus we cannot localize an event with individually contributed data. To this end, we infer the event location from the overlapping attention area using data contributed by the opportunistic community formed in event sensing.

First, as shown in Fig. 5(a), an attention area in the form of a trapezoid (we call it an attention trapezoid) is created according to the photographing location and shooting direction, where *er* is the location error (e.g, GPS error) and *mvd* denotes the assumed maximum visual distance between an event and a viewer (e.g., 50 meters). $\theta$ represents the shooting direction, which is the angle between the shooting direction and the eastward. To make it simple, we assume that the visual range of the camera is around $\pi/3$ (i.e., $2*\eta=\pi/3$), which is considered in constructing the trapezoid.

Second, to localize an event, we divide the map into same-sized grids (e.g., 5m*5m), and we name the length of a grid as *glen*. The grids covered by an attention trapezoid might be the place where the event is going on. However, considering the implicit HI that people often place the target in the middle of a picture, we derive that each covered grid has different probability to be the event place. We thus assign a location weight to each covered grid. Instead of using even distribution, to characterize the implicit HI in photographing, we use the normal distribution to value each grid's location weight. If the point *R* in Fig. 5(a) denotes the centroid of a grid *g*, $g \in Gr$, the location weight of *g*, denoted by *locWeight(g)*, is calculated by Eq. (1).

$$locWeight(g) = \frac{1}{\sqrt{2\pi}\sigma}e^{-\frac{x^2}{2\sigma^2}} \text{ where } x = \frac{len(RS)}{len(US)} \quad (1)$$

where *len*(.) refers to the length of a line, *RS* and *US* are the lines shown in Fig. 5(a), and σ is used as 0.5 in our work.

Third, depending on a number of attention trapezoids, the cumulated weights of grids are normalized and those grids whose location weights exceed a threshold compose the event location. As shown in Fig. 5(b), given *n* pictures, we can get *n* attention trapezoids (see Fig. 5(b)). The accumulated location weight denoted by *alc* of a grid is the sum of *locWeight*s calculated according to these location trapezoids. After being normalized, *alc* of each grid is between 0 and 1. The grid subset *Gr'* of *Gr* (*Gr'*⊆*Gr*) consists of all grids whose *alc*s are over the threshold *loc_th* and then *Gr'* is the *event location*, whose centroid is the event localization result.

*mvd* is an important parameter in attention trapezoid construction. We may assign it a static value (we call it SMVD), e.g., 45 meters. However, if the size of an object is large enough (e.g. the fire in a tall building), people may take pictures of it from over 45 meters away. If we still use 45 meters, then some attention trapezoids might have no overlapping area. Therefore, an adaptive and dynamic *mvd* is used (we call it DMVD) in our work. Considering that the position of an event might be in the middle of different viewers, DMVD calculates *mvd* by Eq. (2).

$$mvd = \frac{1}{2} * \frac{maxD}{sin\left(\frac{|\theta_i - \theta_j|}{2}\right)} \quad (2)$$

where *maxD* is the maximum geo-distance among the crowd contributed pictures, and $\theta_i$ and $\theta_j$ denote their shooting angle. An example is shown in Fig. 6, where there are four pictures taken by four viewers, and *maxD* is obtained between viewer 1 and viewer 4.

*b) Streaming data segmentation with crowd behaviors*. Instead of using visual content analysis, we segment crowdsourced picture stream into subevents with implicit HI. In particular, through a set of user studies, we identify the temporal patterns of crowd-event interaction, including both individual posting and community contribution patterns. Two lightweight subevent detection methods are then proposed based on these findings.

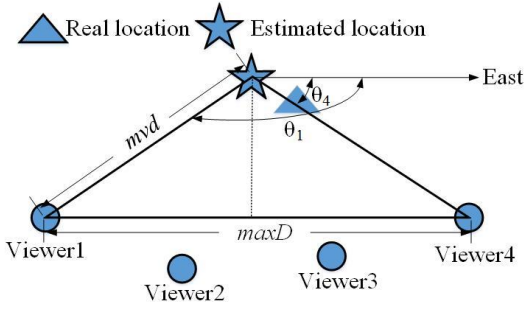

Figure 6. An illustration of DMVD calculation.

Generally, people take pictures of an event when something is interesting or important. To observe the characteristics of moments when people take pictures, we downloaded seven event videos from Youku[8], including one fire-fighting activity, one rescue activity, three street shows, and two talent shows. Eight university students were recruited to tag the moments that they would like to take pictures for these events. To understand the relationship between event evolution and picture taking, we segment the seven events into subevents. Three experts on social event study are asked for event segmentation, where three major factors are considered: *new entity joins* (e.g., an ambulance car arrives), *the entity status changes significantly* (e.g., a person falls down), *the interaction status between a human and an object changes* (e.g., a person leaves the house).

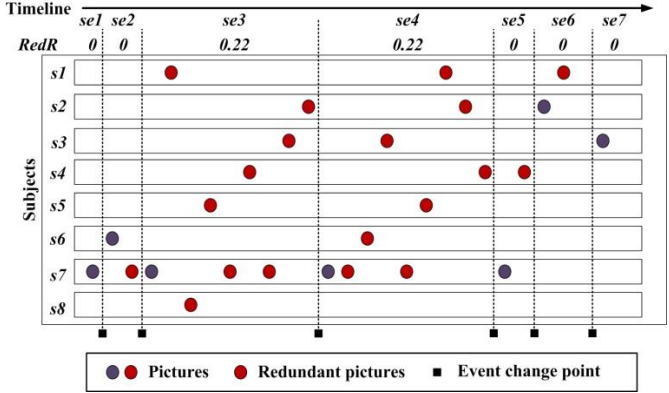

Figure 7. Picture timestamp distribution of the fire-fighting activity.

The temporal distribution of pictures to the fire-fighting activity is shown in Fig. 7. In general, many participants may witness the same event and supply data about it. However, they individually decide what to share and may upload pictures already shared by others, resulting in high redundancy. Therefore, from the results shown in Fig. 7, we find that most subjects would like to take pictures again when the event evolves (we obtain 12 subevents for this activity). However, as bias exists, some people might take more pictures than others when they are interested in what is going on. For each sub-event, if a subject takes two more pictures during it, we consider that the sub-event has redundant submissions. Specifically, we use the redundancy ratio (*RedR*) to reflect the redundancy in the sub-event, which is calculated as the ratio of the number of redundant images to the total number of images taken by people in the sub-event. With this, we can calculate the *RedR* at the *sub-event level* in data collection, i.e., the number of sub-events that has redundant submissions to the whole sub-event set size of an event. The average *RedR* of the subjects to the seven videos is given in Table 2, and the average *RedR* of the subjects to each sub-event of the fire-fighting activity is shown in Fig. 7. We find that the average *RedR* is 20%, which is much smaller than the redundant ratio of the complete picture set (*RRc*, at the *individual picture level*). This observation motivates us to develop implicit HI based event segmentation method.

Table 2. The *RedR* of the seven videos.

|  | *e1* | *e2* | *e3* | *e4* | *e5* | *e6* | *e7* | **Mean** |
|---|---|---|---|---|---|---|---|---|
| *RedR* | 0.30 | 0.09 | 0.25 | 0.06 | 0.13 | 0.28 | 0.29 | 0.2 |
| *RRc* | 0.89 | 0.57 | 0.62 | 0.73 | 0.83 | 0.88 | 0.89 | 0.77 |

Given a picture stream *P* including a series of pictures taken by users, one segment position before a picture $p_m$, and the sub-stream $S_i$ which is the subset of a picture stream. The stream can be segmented again only if either the stream stops increasing or the sub-stream $S_i=\{p_m,...,p_n\}$ meets certain conditions as used in one of the following methods.

- *Crowd-behavior-based segmentation (CS).* According to our findings, if a viewer has been active for an event, she probably takes a new picture when the event changes. Therefore, an interesting moment of the event might be captured by most viewers (i.e., a crowd behavior pattern). Using CS, if the ratio of the number of picture contributors of sub-stream $S_i$ over the total number of nearby viewers (denoted as $\mu(S_i)$) reaches a threshold $r$ ($0<r\leq1$), *P* is split after $p_n$. It is formulated by Eq. (3), which can measure whether picture $p_{n+1}$ belongs to sub-stream $S_i$ or $S_{i+1}$.

$$\begin{cases} p_{n+1} \in S_i, & if\ \mu(S_i) < r \\ p_{n+1} \in S_{i+1}, & if\ \mu(S_i) \geq r \end{cases} \quad (3)$$

- *Crowd-individual-behavior based segmentation (CIS).* CS only considers about crowd behavior, while individual behavior pattern is not considered. As presented earlier, participants tend to take a new picture when the event evolves, which may indicate the appearance of a new sub-event. However, only using individual behaviors may result in segmentation errors, since bias exists and some people may take many pictures within a sub-event that she is interested in. Based on these observations, we combine individual and crowd behavior patterns and propose CIS. In CIS, the CS condition should be met and there should be at least one person who takes more than one picture. Equation (4) is used by CIS to assess which fragment $p_{n+1}$ belongs to.

$$\begin{cases} p_{n+1} \in S_i, & if\ \forall p_t(p_{n+1}.r \neq p_t.r, p_t \in S_i) \vee (\mu(S_i) < r) \\ p_{n+1} \in S_{i+1}, & if\ \exists p_t(p_{n+1}.r = p_t.r, p_t \in S_i) \wedge (\mu(S_i) \geq r) \end{cases} \quad (4)$$

Let us use an example to explain the above segmentation rules. Assuming that pictures $\{p_1,..., p_7\}$ are orderly taken by viewers $\{v_1, v_3, v_5, v_2, v_5, v_6, v_5\}$ respectively. If $r=0.5$, the segmentation is $\{\{p_1, p_2, p_3\}, \{p_4, p_5, p_6\}, \{p_7\}\}$ by using CS, and $\{\{p_1, p_2, p_3, p_4\}, \{p_5, p_6, p_7\}\}$ by CIS.

[8] www.youku.com

## C. Evaluation

This section presents the experiments to CrowdEvent to demonstrate the effectiveness of using implicit HI for crowd event sensing.

*(1) The dataset*. A real-world collected dataset is used for testing implicit HI-based event localization. Both the datasets collected from online-tagging and offline-photographing are used for event segmentation.

*Location dataset*. To mimic the location of events, we collected pictures for 10 outdoor objects (e.g., a building, a monument) in our campus. To obtain diverse samples and evaluate the impacts of geographical distribution of these samples on event localization accuracy, we selected different-sized targets, including a 5-meter-high sculpture, a 20-meter-high library, and a car in the parking lot. For each object, we collected 5~10 pictures.

*Event dataset with online tagging*. We collected 21 social events from Youku, topics including street performance, talent show, accidents, talks, etc. The length ranged from 20 seconds to 9 minutes. 15 subjects were recruited for evaluation. A video-tagging tool was developed for mimicking the photographing behavior based on video viewing. Supposing that each subject was on the spot of the events, they were asked to "take pictures" by pressing the *photographing* button of the tool while watching the event videos. The corresponding frame was regarded as the picture taken. We collected a total of 766 pictures for the 21 social events, an average of 36 for each.

*Event dataset with offline photographing*. Besides online tagging, we recruited subjects to collect data for two real-world events in our campus. They are all about presentation events. 152 and 176 event pictures were collected for the two presentations by 9 and 12 viewers, respectively.

Still, we asked the three experts to segment the events according to the event segmentation rules, which are used as the ground truth for performance measurement. After event segmentation with experts, we find that the picture redundancy ratio to each event ranges from 54% to 88%.

*(2) Baselines and metrics*. We define different metrics to evaluate the effectiveness of our approach.

*Event localization*. To evaluate the performance of event localization, we calculate the Euclidian distance of the detected location with the ground truth, denoted by *e*. The localization method in iSee [40] is used for comparison. iSee locates physical events by leveraging the crowd swiping information on their smart phone's touchscreen in the direction of the event. However, it uses a static *mvd* value 45 meters (SMVD) for localization, and the probability-based grid weighting scheme is also not used.

*Event segmentation*. To evaluate the quality of segmentation, we use the pair-counting *F*-Measure which is often used in data clustering [41]. Supposing that the computed partitioning is *S* and the ground truth partitioning is *G*. $Pairs_S$ and $Pairs_G$ denote the pairs of each partitioning of *S* and *G*, respectively. If we assume that $S=\{\{1,2\},\{3,4,5\}\}$, $Pairs_S=\{(1,2),(3,4),(3,5),(4,5)\}$. Using $Pairs_S$ and $Pairs_G$ the precision and recall of the segmentation are computed by (5) and (6), respectively.

$$Precision\_S = \frac{|Pairs_S \cap Pairs_G|}{|Pairs_S|} \quad (5)$$

$$Recall\_S = \frac{|Pairs_S \cap Pairs_G|}{|Pairs_G|} \quad (6)$$

We introduced two baselines for comparison: namely *mean segmentation* (MEAN) and *similarity-based segmentation* (SIM). Given a constant *K* (which is the number of ground-truth segmentations), the entire picture set is equally split into *K* fragments by Mean. To make comparison, we set *K* to be the same value as the number of segmented sub-events by CIS, i.e., *K*=|*CIS*|. SIM partitions the event according to image similarity measured by the Euclidean distance of images' GIST [42] and color histogram (HIST) [43] features. Similarly, for the purpose of comparison, we adjust the similarity threshold of SIM to make it split the picture stream into *K* segments. If the similarity of two subsequent pictures is less than the trained similarity threshold, the picture stream is segmented at that point by SIM.

*(3) Experimental results*. For event localization, four different *glen*s (grid length) are used and the experimental results are shown in Fig. 8. From the result it is clear that the localization error of our method is much lower than iSee's. It is mainly because that we use dynamic *mvd* and probability-based grid weighting (inspired by implicit HI).

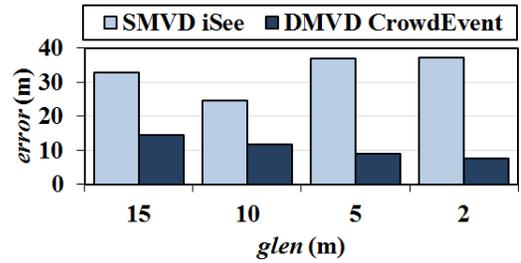

Figure 8. Event localization error.

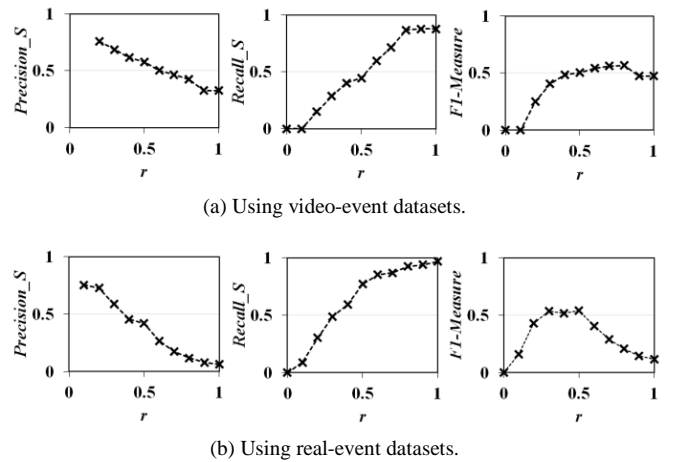

(a) Using video-event datasets.

(b) Using real-event datasets.

Figure 9. The impact of *r* on performance of CS-based segmentation.

For event segmentation, we first study how the value of *r* affects the performance of CS. Based on the results shown in Fig. 9, we set *r* =0.4 when *F*1-Measure values are the best for both video-event and real-event datasets. Figure 10 compares the performance of four segmentation methods: *CS*, *CIS*, *SIM*, and *Mean*. *CIS* outperforms *CS* and *MEAN* as it considers about both crowd and individual behavior patterns. As *SIM* is based on image processing, the computational complexity of *SIM* is much higher than *CIS*.

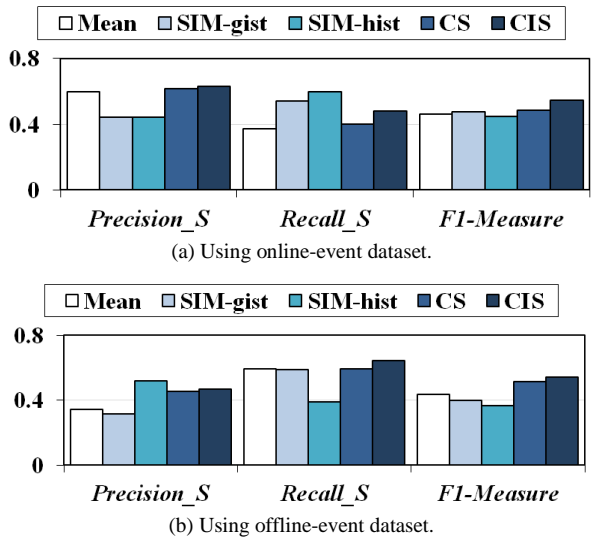

Figure 10. Comparison of segmentation methods by pair-counting: $r = 0.4$ and $K=|CIS|$.

## V. CROWDMINING ON THE WAY WITH CASE STUDY-2: CROWDTRIP

### A. The Scenario and Data-specific Tasks

When paying a visit to an unknown/unfamiliar city, trip planning can be an important, yet time-consuming travel preparation activity. To facilitate the preparation process, numerous studies have been done to recommend a POI (point of Interest) or a sequence of POIs to satisfy users' needs, using the data from social media. However, the issue of planning the detailed travel routes between POIs is ignored, leaving the task to online map services. Such route planning services are mainly based on shortest travel distance or time suggestion, which cannot meet the diverse needs of users. For instance, in the case of travel by driving for a leisure purpose, the scenic view along the travel route would be fun to users. In such contexts, users probably are not rushing to reach the next destination, and the visual and scenic attributes along the driving route are more desirable. Thus, without emphasizing the shortest distance, we aim to recommend the travel route between two POIs with the best scenic view for visitors under a given distance budget.

There are many challenges to be addressed in this case study. Here we mainly focus on two of the following issues.

*a) Scenic Road Network Modeling*. The goal of scenic road network modeling is to score each road segment in the road network according to its nearby scenic information, which is a basis for best scenic view route generation. Nevertheless, road segment scoring is rather subjective and can be impacted by several factors, such as its beauty, popularity, and user preference. This makes the modeling task quite challenging.

*b) Travel route generation*. Another issue in trip planning is about travel route generation, i.e., to select and connect the scenic road segments with high score under given constraints (e.g., the travel distance budget, the starting and ending points). This problem is usually NP-hard, and heuristic algorithms are often employed.

### B. The Usage of Implicit HI

We develop CrowdTrip and use implicit HI to address the two issues above.

*a) Scenic road network modeling with implicit HI.* To measure the scenic beauty of a road segment, we consider two implicit HIs that can be extracted from social media. First, POIs usually have a high density of surrounding geo-tagged photos from social sharing sites (e.g., Flickr[9]). Second, popular roads where users can glimpse more natural view or road-side tourist attractions (e.g., churches, palaces, squares) are also preferred during driving. The popularity of POIs, however, can be learned from the check-in data from location-based social networks (e.g., FourSquare[10]). In this way, we make use of the complementary information provided by both geo-tagged images and check-in data to score the scenic view of a given road segment.

- *Beauty estimation with geo-tagged photos*. A road segment should be scored higher if it is surrounded by more geo-tagged photos. Given road segmentations of a city denoted by *RS*, and photos of scenic views denoted by *PSV*, we compute the scenic score denoted by *SP* of a road segment $rs_i \in RS$ as Eq. (7).

$$SP(rs_i) = \log |\{psv | dis(psv.loc, rs_i) < \delta\}| \quad (7)$$

where $dis(psv.loc, rs_i)$ computes the shortest distance from the photographing location of the photo $psv \in PSV$ denoted by *psv.loc* to the road segment $rs_i$, and $\delta$ is a user-specified threshold. Only the geo-tagged photos with the distance less than $\delta$ are counted when calculating the density to ensure the visibility.

Table 3. Three groups of POIs.

|   | Group Name | Category Labels |
|---|---|---|
| 1 | Natural scenery | Park, garden, lake, forest, mountain, beach, sea, river, bridge, harbor, scenic, hiking. |
| 2 | Tourist attraction | Museum, palace, church, gallery, memorial, monument, square, zoo, university, historic site. |
| 3 | Others | Restaurant, cafe, hotel, and etc. |

- *Popularity estimation with POI check-ins*. Popular roads where users can glimpse more natural view or road-side tourist attractions should also be scored high. Thus, to score the scenic view of a road, the POIs on or near the road should also be taken into consideration. Inspired by the idea that POIs with some specific categories would contribute relatively more to its scenic view [44], we intentionally divide the POIs into three groups according to their category labels, as shown in Table 3. The check-ins are weighted differently according to the group of the corresponding checked-in venues, which investigates the scenic view and the surrounding POI categories of a travel route quantitatively. Given a check-in dataset denoted by $CK$, each check-in $ck_i \in CK$ can be categorized according to the category of its POI and its category is denoted by $ck_i.cate$. The scenic score of a road segment $rs_i$ denoted by $SC$ can be computed by Eq. (8).

---
[9] http://www.flickr.com
[10] https://foursquare.com

$$SC(rs_i) = \frac{\sum_{k=1}^{3} w_k \cdot |\{ck|dis(ck.loc, rs_i) < \delta \wedge ck.cate = k\}|}{|CK|} \quad (8)$$

where the location of a check-in $ck \in CK$ is denoted by $ck.loc$, $k$ denotes one of the three POI categories shown in Table 3, and $w_k$ is the weight of the corresponding POI category. Through a number of experiments, we empirically set $w_1 = 0.65$, $w_2 = 0.3$, and $w_3 = 0.05$.

The integrated scenic score denoted by $SI$ of a given road segment based on both geo-tagged image and check-in data is given in Eq. (9).

$$SI(rs_i) = SP(rs_i) \times SC(rs_i) \quad (9)$$

*b) Crowdsourced data fusion for travel route generation.* The objective of scenic route planning is to generate a high scored scenic view route while satisfying user-specified constraints, including the starting/ending points of the user and the maximum travel distance. The essence is how to select road segments and determine their order. This problem is NP hard and suffers from combinatorial explosion. On one hand, some road segments have higher scenic view score yet higher travel time cost (i.e., resulting in a longer travel distance). On the other hand, there are many possible orders to travel through the selected road segments, and different visiting order can result in different total travel distance. Furthermore, for most road segments, they have two driving directions. The resulted total travel distance is distinct if the road segment is traversed from different driving directions. An approach with three procedures is proposed, as presented below.

- **Interesting area determination**. We first determine the interesting area according to the locations of the given starting/ending points. The minimal rectangle area which covers the two points is selected as the interested area. Only the road segments lying in the area are qualified to be the candidates for traveling.

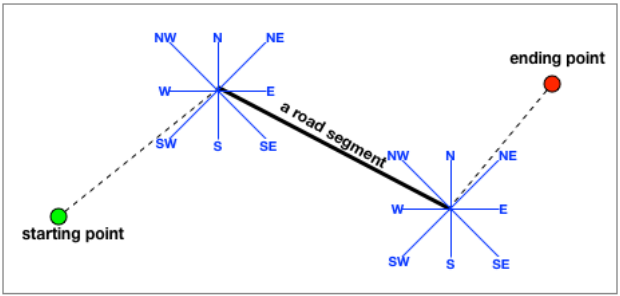

Figure 11. Illustration of driving direction determination for a road segment and a given pair of source and destination.

- *Crowdsourced driving direction determination.* It is hard to determine the driving direction to a road segment. Here, we propose an approach based on implicit HI in human traveling. For a given road segment, we mined the taxi GPS trajectory data to determine its driving direction from the starting point to the ending point. To be more specific, a taxi trajectory is first represented by a sequence of road segments. Then, all the taxi trajectories containing the given road segments are retrieved. Among them, trajectories with sources and destinations *close* to the given pair of starting and ending point are further kept. For a given road segment, we define a trajectory with close source and destination to the given starting and ending points if 1) the direction from the source of the trajectory to the near node of the given road segment is close to the direction from the starting point to that node, and 2) the direction from the near node of the given road segment to the destination of the trajectory is close to the direction from that node to the ending point. An example is shown in Fig. 11. Finally, the frequent driving direction of the kept taxi trajectories is used as the driving direction for the road segment, given the starting/ending points.

- *Scenic route generation*. We have obtained the score for each road segment using crowd intelligence, and the next step is how to select road segments and determine their order to generate a high-score route. We designed three road segment selection strategies. (*i*) To maximize the scenic score, an intuitive idea is to select the road segment with the highest score in the interesting area and add it into the travel route, which is called the Highest-score-first Selecting (*HfS*) strategy. However, this strategy may perform poor if the highest-score road segment is far away, declining adding more high-score road segments. (*ii*) To make a trade-off between the scenic score of an individual road segment and the number of road segments, we propose a heuristic algorithm called Probability-based Selecting (*PbS*) strategy. It is based on the idea that the road segment with a higher scenic view score would be selected and added into the route with a higher probability. (*iii*) As a comparison, the strategy which is based on selecting the road segment randomly regardless its scenic view score is also adopted. We call it the Random-based Selection (*RbS*) strategy. Note that, for *PbS* and *RbS* selection strategies, the algorithm should run repeatedly for a given number of times to ensure a high-score travel route. Each run is independent and can be easily implemented in a parallel way.

A heuristic-based algorithm is proposed, as summarized in *Algorithm* 1. Lines 1~4 refer to the initialization process. Based on user travel route query, the interested area and the driving direction of the road segments in the area will be first determined. We also rank the road segments according to their scenic view score (Line 2). The travel route is initialized as the shortest path from the starting point (i.e. $po$) to the ending point (i.e. $pd$) (Line 3). The shortest path is computed based on the Dijkstra algorithm [45]. Lines 4~7 illustrate the iterative process of road segment selection and ordering. At each iteration, a new road segment will be selected based on *HfS*, *PbS* or *RbS*, and then added in the current travel route (Line 6). The road segment selection and route extension methods are discussed later. The iteration will be terminated once no more new road segments can be added under the maximum distance constraint (Line 5).

Another key issue is route extension by adding new selected road segments. Based on the proposed selection strategies, assume that a road segment *(a1->a2)* is selected and added into the travel route *TR*. The shortest path algorithm is used to link the starting/ending points and the selected travel routes. At the second iteration, a new road segment *(a3->a4)* would be selected, and there are two ordering options to insert it (*a1->a2->a3->a4* or *a3->a4->a1->a2*). We, however, will choose the one with the least distance as the new route. More generally, there are (*n*+1) options that the newly selected road segment can be added if the travel route has already added *n* road segments in.

**Algorithm 1** Algorithm of scenic travel route planning
**Input**: Maximum travel distance constraint *distmax*.
  Starting point *po*;
  Ending point *pd*;
  Scenic Road Network *GS*.
**Output**: A travel route (*TR*)
1: $G'$=GetInterestedArea (*GS*, $p_o$, $p_d$); // The determination of the interested area.
2: *RS*=Rank (GetRoadSegment ($G'$) ); //Get and rank all road segments on area $G'$ according to the integrated scenic score;
3: *TR*=shortestPath($p_o$, $p_d$); *RS*=*RS*-*TR*;
4: **While** (*dist*(*TR*)<*distmax*) **do**
5:   *es*=SelectRoadSegment(*RS*);
    *TR*←*TR*+{*es*};
6:   *RS*=*RS*-{*es*};
7: **end while**

### C. Evaluation

This section presents the experiments to CrowdTrip to demonstrate the effectiveness of using implicit HI for crowd trip planning.

*(1) Dataset*. Three datasets in the Bay Area in the city of San Francisco are used, i.e. the road network, the geo-tagged image data, and the check-in data. The statistical information about the three datasets is shown in Table 4.

Table 4. A statistics of the datasets used by CrowdTrip.

| Datasets | Properties | Statistics |
|---|---|---|
| Geo-tagged photos from Flickr | # of images | 31,022 |
|  | # of users | 1,571 |
| Check-in data from FourSquare | # of check-ins | 110,214 |
|  | # of users | 15,680 |
| Road network from OpenStreetMap[11] | # of nodes | 3,771 |
|  | # of road segments | 5,940 |

*(2) Evaluation metrics*. The scenic score of the travel route is used to measure the effectiveness of the proposed travel route algorithm, which is defined as the sum of the scenic score of all road segments that the travel route contains. We formulate it in Eq. (10).

$$\sum_{i=1}^{|TR|} SI(tr_i) \qquad (10)$$

where $tr_i \in TR$, *TR* is the travel route.

---

[11] http://www.openstreetmap.org/

*(3) Experimental results.* We compare our scenic road network modeling approach to the previous work which is solely based on geo-tagged image data [46]. It fails to take category information about POIs into consideration which is rooted by the fact that geo-tagged images do not contain such information explicitly. Taking the two road segments shown in Fig. 12 for example, the road in the left is the "*Bicycle Route 65*", which is a famous driving road regarding to its attractive scenic view. As can be seen, though "*Bicycle Route 65*" is surrounded by only four POIs (i.e. 'Conservatory Flows', 'Gold Gate Park', and so on), it also offers a pleasant driving experience. The road in the right is the "the 7th St", which is a regular and normal commercial urban street. It is surrounded by at least eight popular POIs (twice more than that in "Bicycle Route 65"), but they are more preferred by shoppers. Hence, the road in the left should be scored higher. The "Bicycle Route 65" is ranked 23rd by our modeling approach but ranked 234th by the approach in [46]; "the 7th St" ranked 1,004th by our modeling approach but ranked 12th by approach in [46]. In summary, our modeling approach achieves a more reasonable result as we extract the complementary information provided by two data sources.

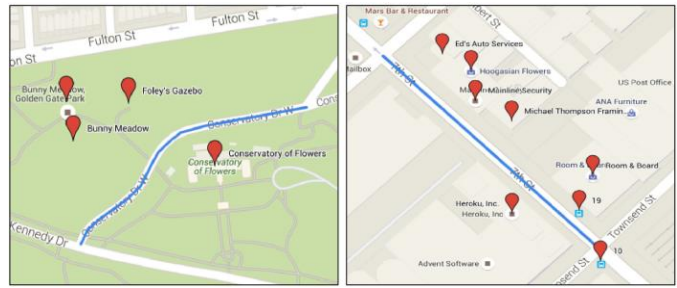

Figure 12. Two road segments: The road in the left is a famous driving road for its attractive scenic view; while the road in the right is a regular commercial urban street.

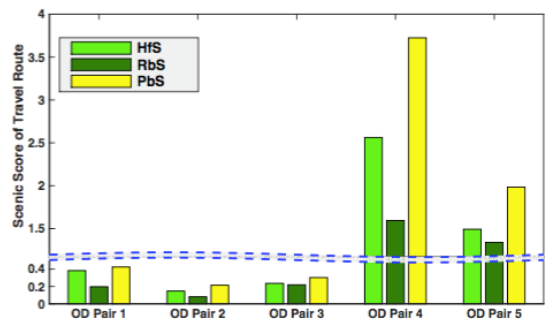

Figure 13. The scenic score of the travel routes resulted by different route planning approaches.

We also conduct experiments to the three route planning methods (i.e., *HfS*, *PbS*, and *RbS*), and present the comparison results in the scenic score of the travel route. In particular, we select five Origin-Destination (OD) pairs (in different areas) for the evaluation.

Figure 13 shows the results of the scenic score of the planned travel routes for the selected OD pairs. Considering that the difference in scenic scores between the first three and the last two OD pairs is quite great, we adjust the values of *y*-axis in order to show them in the same figure. Therefore, the region between the double-dotted line indicates omitted values of *y*-axis (i.e., 0.4-1.5). For all five OD pairs, *PbS* achieves the highest scenic score, while *RbS* results the lowest one. The scenic score obtained by *HfS* is in-between, better than that of *RbS*, worse than that of *PbS*.

## VI. CONCLUSION AND FUTURE VISION

We have presented CrowdMining, which leverages implicit HI for crowdsourced data understanding. The intrinsic concepts and two studies on CrowdMining are presented, which demonstrate how we use implicit HI for solving complex data tasks. A generic model for CrowdMining systems is also proposed. Experiments based on online/offline crowdsourced data indicate that implicit HI can be used effectively to solve various crowd data mining tasks.

This paper reports our efforts to crowdsourced data understanding with the usage of implicit HI. There are several interesting directions to be investigated further in the future, as discussed below.

- ● *The scope of implicit HI.* The major types of implicit HI presented in this paper include crowd behavior patterns, crowd-object interaction patterns, and so on. Implicit HI has wide scope in terms of cognitive abilities, individual attributes, social features, interaction and behavior patterns. It is crucial to characterize them and investigate their usage in crowdsourced data mining.
- ● *The benefits of using implicit HI.* We study the usage of implicit HI for a variety of data mining tasks. In the future it is interesting to extend its application scope and make use of it in other data-centric crowdsourcing tasks and for more-complex problems that are not easy to be solved by pure machine intelligence.
- ● *The emergence of crowd-machine computational systems.* This paper investigates the integration of implicit HI and MI in mobile crowd sensing. Implicit HI is used as feature inputs for machine learning and data mining algorithms. With the manifold efforts of embedding human intelligence in computing systems, we will finally build crowd-machine computation systems. The complementary features of implicit, explicit, and human intelligence should be further explored and new integration/collaboration manners should be studied.